\documentclass[review]{elsarticle}
\usepackage{lineno,hyperref}
\usepackage{color}
\usepackage{longtable}
\usepackage{cleveref}
\usepackage{pdfpages}
\modulolinenumbers[5]

\journal{Journal of \LaTeX\ Templates}

\begin{document}

\begin{frontmatter}

\title{Inconsistencies in {\it ab initio} evaluations of
  non-additive contributions of DNA stacking energies}

\author{Ken Sinkou Qin,Tom Ichibha}
\address{School of Information Science, JAIST}

\author{Kenta Hongo}
\address{Research Center for Advanced Computing Infrastructure, JAIST}
\address{Center for Materials Research by Information Integration,
 Research and Services Division of Materials Data and Integrated System,
 National Institute for Materials Science}
\address{Computational Engineering Applications Unit, RIKEN}
\address{PRESTO, JST}
\author{Ryo Maezono}
\address{School of Information Science, JAIST}
\address{Computational Engineering Applications Unit, RIKEN}


\author[mysecondaryaddress]{JAIST, Asahidai 1-1, Nomi,
  Ishikawa 923-1292, Japan}
\cortext[Ryo Maezono]{Ryo Maezono}
\ead{rmaezono@mac.com}


\begin{abstract}
We evaluated the non-additive contributions
of the inter-molecular interactions in B-DNA stacking by using 
diffusion Monte Carlo methods
with fixed node approximations (FNDMC).
For some base-pair steps, we found that
their non-additive contributions evaluated by FNDMC 
significantly differ from
those by any other {\it ab initio} methods,
while there are no remarkable findings on
their stacking energies themselves.
The apparently unexpected results of non-additivity
raise issues in both FNDMC and correlated wavefunction methods.
For the latter, it can be partly attributed to the imperfect
complete basis set (CBS) correction scheme
due to the limitation of the computational costs. 
On the other hand, the striking contrast between the stacking
and non-additivity behaviors was found in FNDMC.
This might imply that the error cancellations of
the fixed node biases in FNDMC 
work well for the stacking energies,
while not for the non-additivity contributions
involving charge transfers caused by hydrogen bonds
bridging Watson-Crick base pairs. 
\end{abstract}


\end{frontmatter}


\section{Introduction}
The non-additivity in interactions 
is obviously expected in inter-molecular 
bindings due to the induced polarizations 
caused by quantum fluctuations, 
such as vdW (van der Waals) forces. 
Since the binding itself has proven a great 
challenge to {\it ab initio} methods in terms of its 
description and reproduction,
the further issue of non-additivity has prevented
it from becoming a topic of major interest, 
and thus it has not been well analyzed to date.
Most of the current implementations of 
'molecular force fields'
assume the superposition of two-body forces,
applied to a vast number of simulations of 
self-organizations by biomolecules~\cite{sherrill2012energy,
  Bottaroeaar8521,lemkul2017polarizable,
  hostas2015representative,milovanovic2018new,
  hermann2017nanoscale,cole2016biomolecular,riniker2018fixed}.
This assumption has been verified for B-DNA in a previous work~\cite{2006SPO},
reporting stacking energies predicted by highly accurate quantum chemistry methods
are in good agreement with those by 
the empirical force fields such as AMBER~\cite{2003PON}.

\vspace{2mm}
Recent progress in accurate computational methods,
especially diffusion Monte Carlo (DMC), 
has enabled one to apply them to much larger systems~\cite{2002GRO,
2008KOR,2010HON,2012WAT,2013DUB,2013HOR,2014DUB,2013HON,2015HON,2016HON,2016DUB,2017HON,2017ICH}.
FNDMC (fixed node DMC)~\cite{1998KWO} is a widely 
used implementation of DMC, where
an imaginary-time dependent many-body wave function of a system evolves 
and gets closest to the exact (stationary) one by
modifying its amplitude under the assumption that
its nodal surfaces are fixed to be initial ones.
The method is capable of giving
an exact answer provided that the fixed node 
is same as that of the exact solution. 
The fixed nodes are usually generated by 
DFT (density functional theory) or 
MO (molecular orbital) methods, 
which are not generally expected to be exact.
Hence FNDMC inherently includes a systematic bias due to 
the assumption on the fixed nodes (fixed node bias). 
Nevertheless, it has well reproduced binding natures
of weakly-bound systems.~\cite{2010HON,2013HON,2015HON,2014DUB}
For such weakly-bound systems, 
it has been clarified that the non-additivity 
is much greater than was originally expected.~\cite{2015REI,2014MIS}
Supposing non-additive contributions were positive definite,
their effects on molecular stability
would get minimized in order for a
system to be stable because non-additivity
would evidently increase as a molecular size grows.
If so, only a minor correction would be required 
{\it e.g.} for $C_6$ (the coefficient of $1/R^6$
decaying interactions) 
without any qualitative impact.~\cite{2017HON}
This has been demonstrated in
a previous study of B-DNA stacking~\cite{2006SPO},
indicating a pairwise approximation to
molecular potential holds well.
Our main result in the present study
cast doubt on this approximation,
as shown in Fig.~\ref{fig:nonadditive}
where DMC is applied to evaluate the stacking energies of B-DNA 
base pairs~\cite{kilchherr2016single,2015HON,
  1997SPO,2006SPO,2008HIL,2008FIE,2013PAR}
and their non-additive contributions.
While conventional methods including CCSD(T) and most of DFT
predict tiny ($\sim$ several kcal/mol),
positive definite non-additive contributions, 
the DMC does much larger non-additive 
contributions with not only positive,
but also negative signs, depending on base pairs;
in some cases their magnitudes are as large as
the corresponding stacking energies themselves
($\sim$ 10 kcal/mol).
While non-additivity in the conventional methods 
just slightly unstabilizes all the B-DNA base pairs,
that in FNDMC can either reduce (positive sign)
or enhance (negative sign) the stacking interactions
depending on the pairs.

\vspace{2mm}
The negative non-additive contributions can be deduced
from a simple model analysis based on the
London theory~\cite{london1937general} as follows.
Fig.~\ref{fig:structure} (b) shows the schematic geometry 
of Watson-Crick base pairs in B-DNA
specified as 'VW:XY' in the notation convention. 
Base fragment pairs (W,V) and (X,Y) are respectively
located at each of strands (framed by each of red-colored boxes)
to form a complete four-body system.
In the London theory,
dispersion interactions between the upper and lower layers
can be approximated as
$\varepsilon \sim \alpha^{(upper)}\cdot\alpha^{(lower)}$,
where $\alpha^{(upper)}$ and $\alpha^{(lower)}$ are
the dipole polarizabilities of each layer.
Note that the polarizability is roughly additive~\cite{2014BLA}
because it scales as the molecular weight.
This makes a rough estimate of the stacking energy
for the entire (four-body) system as 
$\varepsilon^{(4)}=2\times 2\cdot\varepsilon^{(2)}<0$, 
where $\varepsilon^{(2)}$ denotes the stacking energy 
for a partial (two-body) system.  
(Of course, there are other dependencies of $\varepsilon$  
such as on ionic energies and geometry {\it etc.},
which do not have such a great effect --
our data given in the Supporting Information
have demonstrated that the simplified discussion holds.)
The estimate then gives the non-additivity as 
$\Delta \varepsilon^{(4)} = \varepsilon^{(4)}-4\times
\varepsilon^{(2)} = 0$.
This would be true only for the limit, $l\to 0$ [$(a'/a) \to 1$].
In practical cases [$(a'/a) > 1$], however, we can 
ignore ``cross'' interactions 
[between W and Y and between X and V
  (see Fig.~\ref{fig:structure})]
due to $(1/a^6) \gg (1/a'^6)$ ($(a'/a)$ = 1.5$\sim$1.9).
We then arrive at the negative non-additivity:
$\Delta \varepsilon^{(4)} = \varepsilon^{(4)}-2\times\varepsilon^{(2)} 
= 2\times\varepsilon^{(2)} < 0$.
Upon applying the simplified model analysis,
the London theory produced a negative sign of
the non-additivity contribution,
implying that the stacking energy is {\it enhanced}
by the non-additivity.

\vspace{2mm}
It is worth noting that the negative non-additivity
found in FNDMC and the above-mentioned model analysis
is quite different from the results obtained using 
other {\it ab initio} methods.
In this paper, we provide several possible 
theories explaining why discrepancies arise 
between the FNDMC and the other predictions.
We first investigated that
'negative values are predicted only by DMC'.
Once we verified this issue, 
a further doubt occurs regarding the sign alternation.
We also provide plausible explanations 
for the sign of non-additivity. 

\vspace{2mm}
Our central findings in the present study
are summarized as follows:
Our evaluation of binding energies
come up to {\it common expectations}
for the methodologies (Fig.~\ref{fig:stacking}).
But this appears not the case for
non-additivity at first glance (Fig.~\ref{fig:nonadditive}):
(1) FNDMC and ``CCSD(T)'' results are
not in accordance with each other
[here we use ``CCSD(T)" instead of CCSD(T) 
for some reason described below].
(2) ``CCSD(T)'' gives almost the
same results as B3LYP and Hartree-Fock (HF), 
which hardly happen when evaluating the binding energies.
These {\it apparently} strange conclusions
are likely to mislead readers,
so we shall outline the contents below.
We claim that the conclusions are not
due to careless choices of our
computational specifications,
but more fundamental points --
practical approximations adopted
in DMC and CCSD(T),
each of which works well for
evaluating binding energy,
but not for non-additivity.
In the former case, we can see many previous works
reporting the cancellation of the fixed node biases
between a whole system and its constituent molecules
works well for evaluating 
binding energies~\cite{2013HON,2015HON,2014DUB}.
On the other hand, such a cancellation has not
been investigated yet.
In the latter case, we note the fact that
``CCSD(T)'' applied to B-DNA systems is actually
``CCSD(T) with CBS at the MP2 level''~\cite{2006SPO}.
We finally conclude this practical approximation
can be attributed to the reason why it gives
the same trends in non-additivity as B3LYP
that is not believed to be capable of
reproducing vdW interactions~\cite{2013HON}. 

\section{System and methods}
\begin{figure}[!hbtp]
\begin{center}
  \includegraphics[scale=0.4]{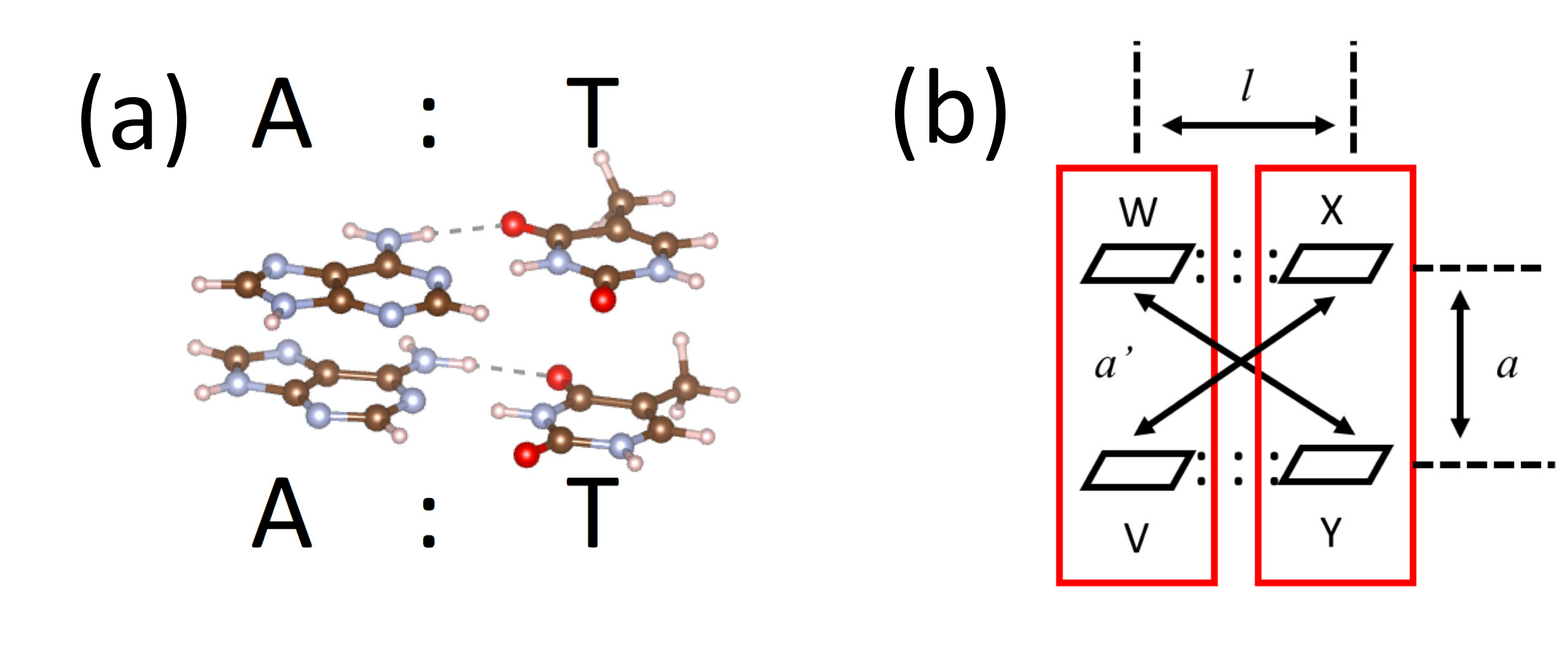}  
\end{center}
~\caption{\label{fig:structure}
    Panel (a) shows an example of the geometry 
    for the 'AA:TT' pair. 
    The notational convention, 'VW:XY', 
    conforms to the standard ~\cite{2006SPO} 
    used in this field, as explained in panel (b), 
    where the bases {V,W,X,Y} appear in this order 
    along $\cap$-shape wise. 
}
\end{figure}

Our target systems are ten unique B-DNA (Watson-Crick) 
base-pair steps shown in Fig.~\ref{fig:structure}.
Their molecular geometries were obtained from
a pioneering work due to \v{S}poner {\it et al.}~\cite{2006SPO},
and they are fixed throughout our simulations.
Note that, {\it e.g.}, 'AT:AT' and 'TA:TA' seem identical
schematically because of their mirror image relation.
Actually, they have different geometries in B-DNA sequences,
and hence they are not identical.
\begin{figure*}[!hbtp]
\begin{center}
  \includegraphics[scale=0.35]{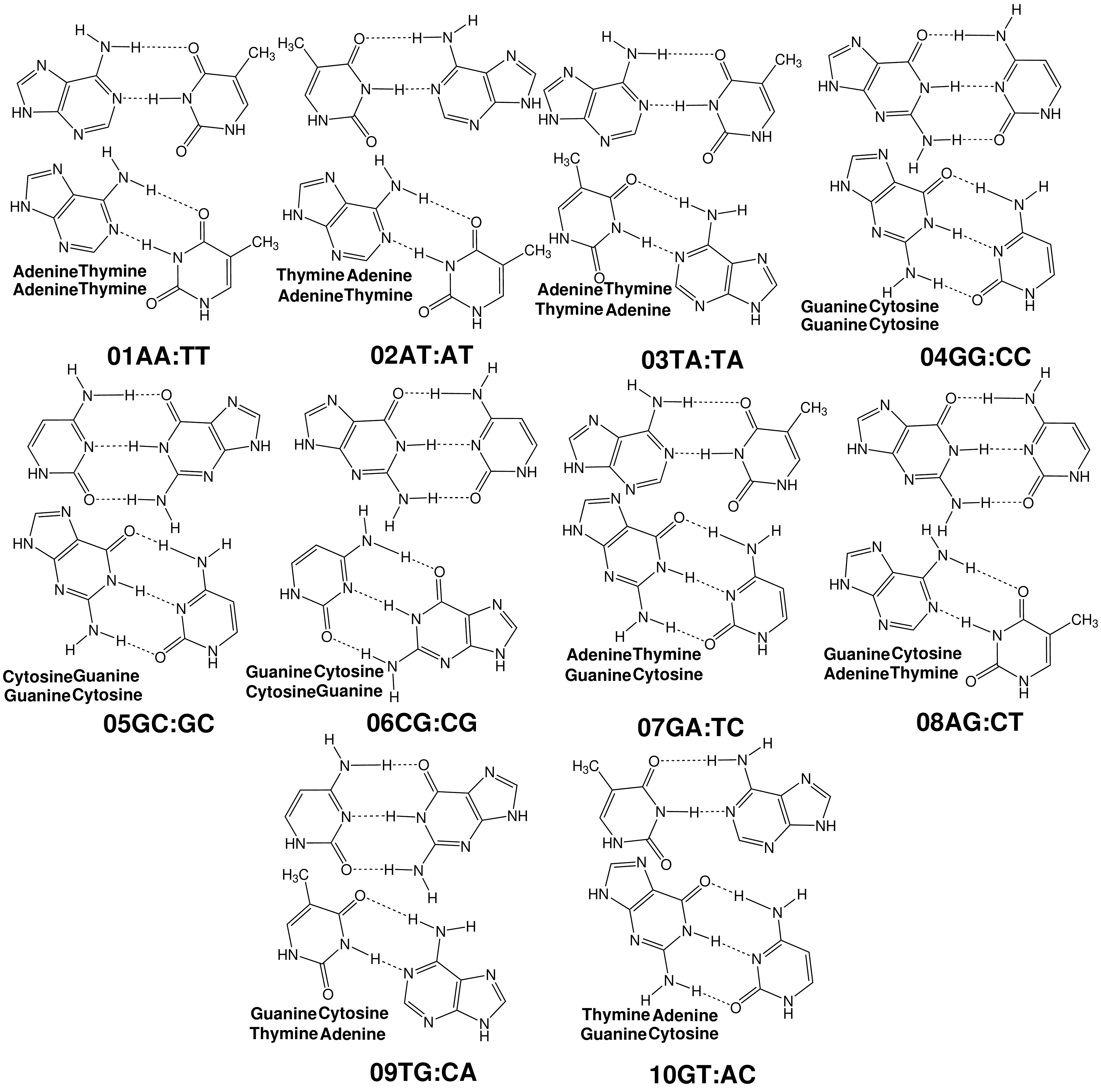}
\end{center}
~\caption{\label{fig:structurec}
    Ten Watson-Crick base pairs 
    Evaluated in B-DNA.
    Each system is composed of four kinds of bases, 
adenine (A), thymine (T), 
guanine (G), and cytosine (C) molecules. 
}

\end{figure*}
\noindent
Stacking energies, $\varepsilon^{(4)}$ and 
$\varepsilon^{(2)}$ are evaluated using {\it ab initio} methods.
In the present study we adopted 
the same non-additive contribution $\Delta \varepsilon^{(4)}$
as that defined by \v{S}poner {\it et al.}~\cite{2006SPO}:
\begin{equation}
  \label{nonadditive}
  \Delta \varepsilon^{(4)} =\! \varepsilon^{(4)}_\mathrm{VW:XY}\! -\! \left(  
  \varepsilon^{(2)}_\mathrm{VW}\! +\! \varepsilon^{(2)}_\mathrm{YX}\! + \!
  \varepsilon^{(2)}_\mathrm{VX}\! +\! \varepsilon^{(2)}_\mathrm{YW} \right) 
\end{equation}
Note that this definition does not include
the contributions of the hydrogen bonding
between bases in a Watson-Crick pair within each step:
V:Y and W:X.

\vspace{2mm}
The dispersion interaction -- a main contribution to stacking --
can be reproduced by going beyond MP2 (M{\o}ller-Plesset) level 
treatment of electron
correlations~\cite{2014MIS,1997SPO,1994CHA,2000CHA}.
We hence applied FNDMC (diffusion Monte Carlo
with the fixed-node approximation)~\cite{1982REY}
to compute the $\varepsilon^{(4/2)}$ values
in Eq.~(\ref{nonadditive})
using the CASINO code~\cite{2010NEE}.
Computational details in our FNDMC were the same as
those in our previous studies~\cite{2013HON,2016HON}.
Furthermore, they met 
a protocol developed for non-covalent systems
by Dubecky {\it et al.}~\cite{2014PCCPDubeck}:
We adopted Burukatzki-Filippi-Dolg pseudopotentials
(BFD-PPs)~\cite{2007BUR} and applied the T-move 
scheme~\cite{2006CAS} to their evaluation
with the locality approximation~\cite{1991MIT};
fixed-node guiding functions were constructed
from DFT-B3LYP/VTZ orbitals with Gaussian09~\cite{2009FRI};
one- and two-body Jastrow functions were considered
where their parameters were optimized
by variance minimization procedure~\cite{2005UMR,2005DRU};
a time step is set to be $\delta \tau = 0.005$, which
is small enough to avoid the time step error bias.~\cite{1993UMR}
\par
A number of previous works have demonstrated that FNDMC
with similar conditions are 
computationally feasible for non-covalent systems as large as
the present B-DNA systems and successfully reproduce
their non-covalent interactions comparable to
CCSD(T)/CBS (coupled-cluster approach at the singles
and doubles level augmented with
perturbative triples using
complete basis set limit).~\cite{2002GRO,
  2008KOR,2010HON,2012WAT,2013DUB,2013HOR,
  2014DUB,2013HON,2015HON,2016HON,2016DUB,2017HON,2017ICH} 
In particular, 
the B-DNA stacking energies evaluated by FNDMC were
calibrated in detail in our previous works.~\cite{2013HON,2016HON} 
  
\par
There is no doubt that CCSD(T)/CBS is 
a state-of-the-art or ``gold standard'' quantum chemistry method
-- an established protocol of capturing dispersion interactions
in non-covalent systems.~\cite{2013REZ}
Its applicability is, however, quite limited to small systems.
Actually, CCSD(T)/CBS is not applicable to
our target systems of B-DNA base-pair steps,
even using the best supercomputer.
The practically best possible solution~\cite{2006SPO} was 
to apply ``CCSD(T)/CBS'' to all pairwise stacking and
add a many-body correction at MP2/VDZ level to the pairwise sum.
Here we note that ``CCSD(T)/CBS'' for the pairs is 
not a {\it true} one, but an approximation such that
MP2/CBS is combined with a energy difference between CCSD(T)
and MP2 obtained using a small basis set.
Hereafter we refer this sort of approximation to CCSD(T)/CBS(MP2).
Very recently, Kraus {\it et al.}~\cite{2019KRU} have attempted to
avoid the approximation of stacking interaction such
as the sum of four base-base stacking energies,
but their level of theory has not reached CCSD(T)/CBS.
Parker {\it et al.}~\cite{2013PAR} stated that 
such an approximate estimate can be used for reference, 
but not as a conclusive standard value.
In the above-mentioned context, a {\it true} non-additivity
at CCSD(T)/CBS level of theory remains unknown.
\par
In addition to the 'correlation-level' non-additivity,
we also investigated 'SCF-level' non-additivity by performing
Hatree-Fock (HF) and DFT with various choice of 
exchange-correlation (XC) functionals. 
Our HF and DFT simulations were carried out using
Gaussian09~\cite{2009FRI}
with the same basis set and pseudo potential as FNDMC.
It is well known that some of our choices of XC functionals here
({\it e.g.}, LDA and B3LYP) 
are inappropriate for describing 
the interactions in the present system, 
but we intentionally adopted them for comparison
between correlation- and SCF-level non-additive contributions.
In addition to the conventional XC functionals, we adopted
recently developed XC functionals such
as $\omega$B97X, $\omega$B97M-V~\cite{2011SIN},
B3LYP-D3~\cite{2010GRI}, and CAM-B3LYP-D3~\cite{2004YAN}
in order to investigate not only dispersion effects,
but also hydrogen bonding.
\par

\section{Results}
The present study adopts CCSD(T)/CBS[MP2] results
due to \v{S}poner {\it et al}~\cite{2006SPO} as reference,
though they are {\it not} ``true'' CCSD(T)/CBS
as described in the previous section ``System and methods''.
Here we show not only non-additive contributions
$\Delta \varepsilon^{(4)}$, but also stacking
energies $\varepsilon^{(4)}$ themselves similar
to conventional researches.
We will see that FNDMC gives rise to 
a striking dependence of $\Delta \varepsilon^{(4)}$
on base-pair steps, compared to the other
{\it ab initio} methods including CCSD(T)/CBS[MP2]
as well as DFT, while there is not any novel findings
in their $\varepsilon^{(4)}$ evaluations.

\subsection{Stacking energies $\varepsilon^{(4)}$}
\begin{figure*}[hbtp]
\begin{center}
  \includegraphics[scale=0.75]{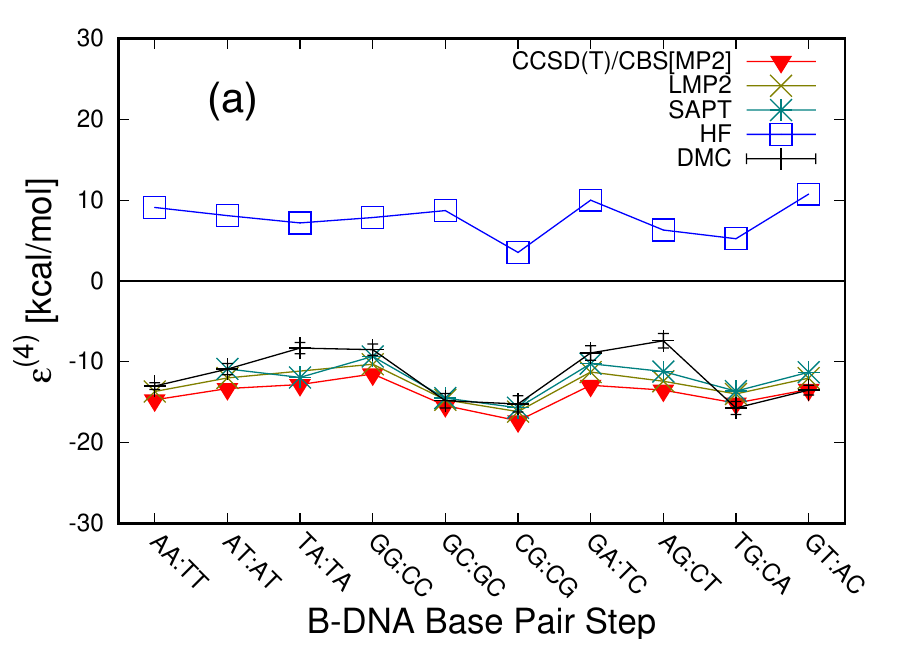}
  \includegraphics[scale=0.75]{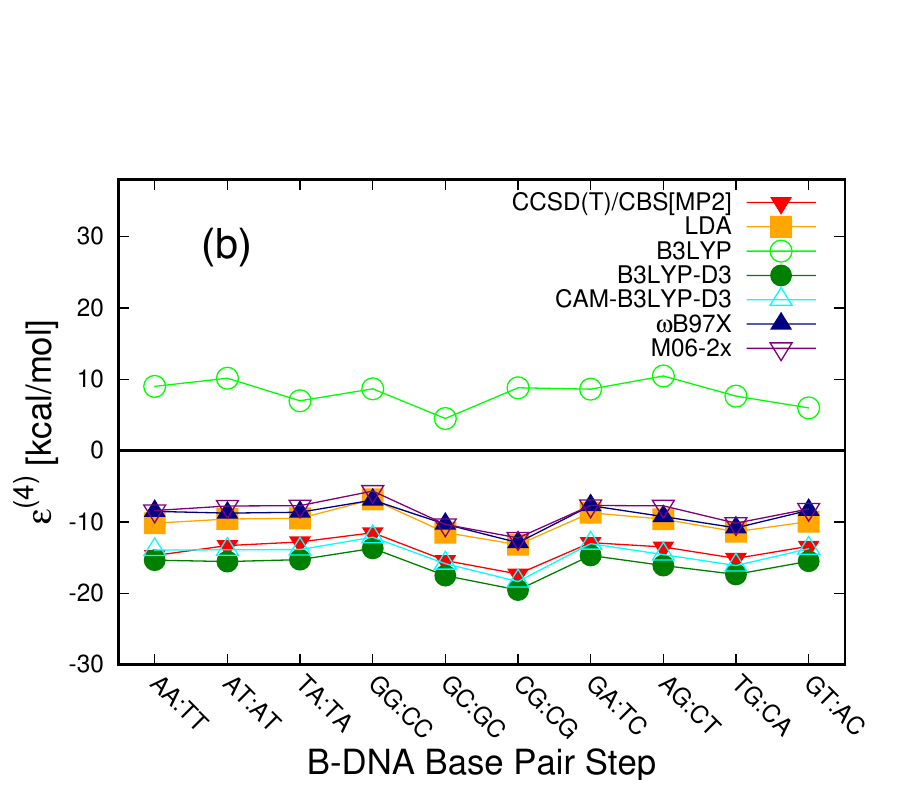}
\end{center}
~\caption{\label{fig:stacking}
    Four-body stacking energies, $\varepsilon^{(4)}$ [kcal/mol], 
    for the B-DNA base-pair steps evaluated by various methods.     
    The negative values correspond to the binding, 
    and hence we see that only B3LYP cannot properly describe 
    the binding. 
    CCSD(T) values were taken from a previous work~\cite{2006SPO}.
}
\end{figure*}
We shall start with the four-body stacking energies ($\varepsilon^{(4)}$) of
the B-DNA base-pair steps. Fig.~\ref{fig:stacking} (a) and (b)
are the $\varepsilon^{(4)}$ values obtained from wavefunction-based
and DFT-based methods, respectively, together with CCSD(T)/CBS[MP2]
as reference. Within the wavefunction methods (Fig.~\ref{fig:stacking} (a)),
we can see that HF fail to describe the stacking properly because it does not
include electron correlations at all by nature. It is obvious that
an appropriate description of stacking requires
theory more than MP2 level~\cite{1994CHA,2000CHA}.
Overall, the correlated methods agree with each other. Looking closely at the trend,
CCSD(T)/CBS[MP2] was found to overbind compared with the other approaches,
LMP2, SAPT and FNDMC. This can be attributed to two practical approximations adopted
in the CCSD(T) reference~\cite{2006SPO}: Since the B-DNA system is too large
to be computed by CCSD(T) with larger basis sets (cc-pVTZ and cc-pVQZ),
the {\it true} CCSD(T)/CBS is not available in any literature so far.
Instead, two approximations were applied to estimate ``CCSD(T)/CBS''.~\cite{2006SPO}
Firstly, all the pairwise base-base terms ($\varepsilon^{(2)}$ in Eq.~(\ref{nonadditive})),
were computed by MP2/CBS plus an energy difference between CCSD(T)
and MP2 with a common small basis set (6-31G*(0.25)).
Secondly, the non-additive contribution ($\Delta \varepsilon^{(4)}$)
was evaluated at RI-MP2/aug-cc-pVDZ level.
In this sense we refer the CCSD(T) to ``CCSD(T)/CBS[MP2]''.
From the above facts we can infer that ``CCSD(T)/CBS[MP2]''
overbinds, similar to MP2.
Note that both SAPT and DF-LMP2 are known to correct the overbinding
trend in MP2~\cite{2008HIL}.
Accordingly, 'the overbinding in CCSD(T)/CBS[MP2]' can be also supported
by the fact that the magnitudes of stacking energies in SAPT and DF-LMP2
are smaller than those in CCSD(T)/CBS[MP2], as can be seen in Fig.~\ref{fig:stacking}
(a). It was found that FNDMC deviates from the other correlated methods
depending on base-base pair steps (especially TA:TA and AG:CT),
which is closely related to a significant difference in non-additive
contributions between the methodologies, as described later.
\par
As for the DFT methods, we see that only B3LYP cannot properly describe
the stacking, which is consistent with its well-known deficiency in
describing dispersion interactions~\cite{2012COH}.
On the other hand, recently developed dispersion methods within DFT
reproduce the stacking for all the steps~\cite{2010GRI},
giving almost the same trend as CCSD(T)/CBS[MP2].
When comparing with CCSD(T)/CBS[MP2], B3LYP-D3
(B3LYP with an empirical dispersion correction) gives the
wiggling stacking energies; CAM-B3LYP-D3 (B3LYP-D3
with the long-range corrected (LC) exchange) slightly
overbinds overall; M06-2X (hybrid meta GGA) and
$\omega$B97X (B97 functional with long- and short-range
corrected exchange) both underestimate the stacking energies 
for all the steps, which has been also demonstrated for
other non-covalent systems~\cite{2015HON}.
Note that the stacking described by LDA is known
to be artificial.~\cite{2010HON,2012WAT,2013HON,2015HON,2016HON,2017HON,2004HAS,2008TKA}

\subsection{Non-additivity $\Delta \varepsilon^{(4)}$}
\begin{figure}[!hbtp]
\begin{center}
  \includegraphics[scale=0.75]{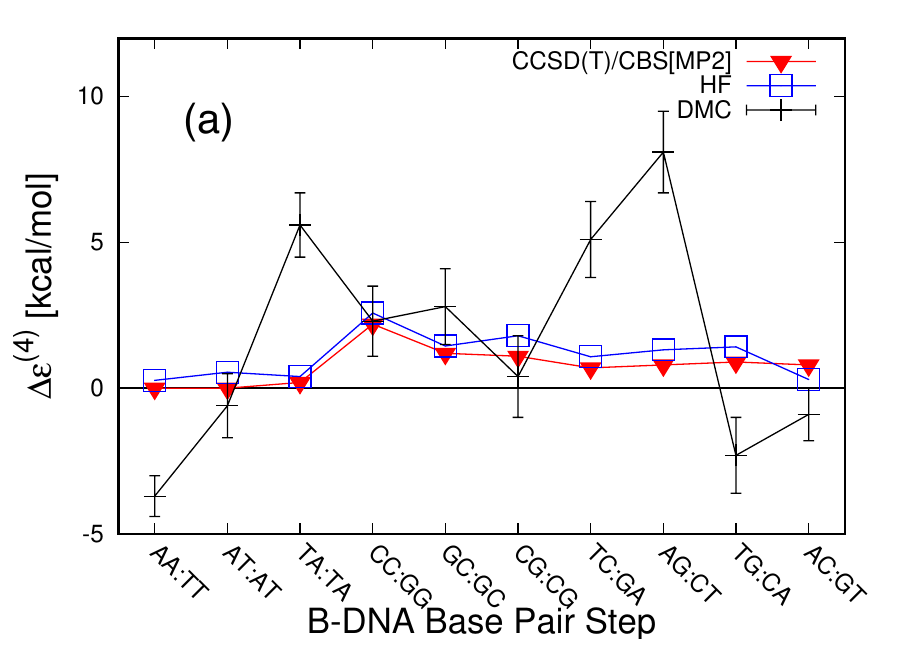}
  \includegraphics[scale=0.75]{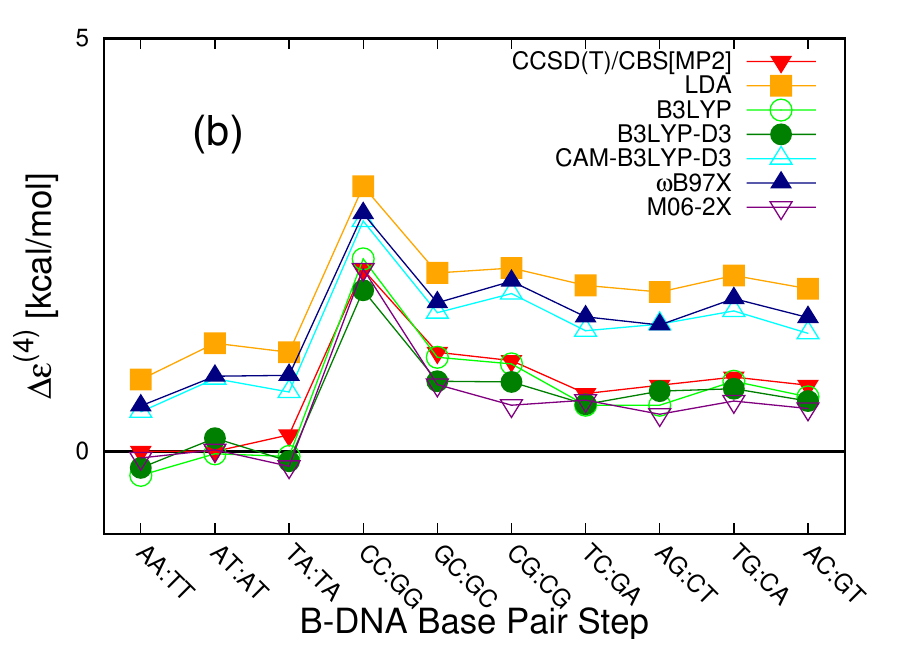}
\end{center}
\caption{\label{fig:nonadditive}
    Non-additive contribution, $\Delta E^{(4)}$ [kcal/mol], 
    evaluated by various methods. 
    DF-LMP2+$\Delta$(T)~\cite{2008HIL} and 
    SAPT0/jaDZl~\cite{2013PAR} appearing in Fig.~\ref{fig:stacking}
    are not shown here because their non-additive contributions are 
    not available. 
    For CCSD(T), the data is taken from the preceding 
    work.~\cite{2006SPO}
    Unlike stacking energies (Fig.~\ref{fig:stacking}), 
    CCSD(T) agrees with both HF and B3LYP, while
    it is far from FNMDC.
    Plausible discussions for this are given in the text. 
}
\end{figure}

Non-additive contributions $\Delta \varepsilon^{(4)}$
evaluated from various methods are shown in
Fig.~\ref{fig:nonadditive}.
We found FNDMC giving a wiggling behavior of $\Delta \varepsilon^{(4)}$,
compared to all the other methods.
This remarkable sign alternation found in FNDMC is a
central issue in the present study.
\par
We first remark that the non-additive contributions do not necessarily
arise from the electron correlations but always appear as non-linear
processes inside a many-body system.
In terms of the perturbation expansion based on SAPT,
an interaction energy is decomposed into physically meaningful components:
electrostatic, induction, exchange, and dispersion terms.~\cite{1994CHA,2000CHA}
Even at the Hartree-Fock (HF) level of theory, the 'exchange' and 'induction'
parts of the non-additivity occur~\cite{1994CHA,2000CHA}, which we refer to
'SCF-level non-additivity' hereafter.
The behavior of HF in Fig.~\ref{fig:nonadditive}
can be an appropriate reference to the 'SCF-level non-additive' contribution
because HF is incapable of describing the dispersion by nature.
Except for FNDMC, CCSD(T)/CBS[MP2] as well as all the other DFT methods
exhibit almost the same $\Delta \varepsilon^{(4)}$ as HF;
their magnitudes of $\Delta \varepsilon^{(4)}$ are slightly
increased or decreased from those of HF owing to their balance
between exchange and correlation (as described later),
but their differences are quite small (less than 1 kcal/mol).
Presumably, the non-additive contributions appearing inn all the DFTs
and even CCSD(T)/CBS[MP2] may be regarded as being 'SCF-level' ones
or hardly describing 'dispersion-level/correlation-level' ones.
On the other hand we may imply that FNDMC describes more dispersion/correlation
contributions than the other methods.
\par
Looking closely at individual methods, it was found from Fig.~\ref{fig:nonadditive} (a)
that the behavior of CCSD(T)/CBS[MP2] is the most striking:
The electron correlation described by CCSD(T)/CBS[MP2] hardly makes any corrections
to the SCF-level non-additivity. This means that the 'correlation-level non-additivity'
cannot be well captured by CCSD(T)/CBS[MP2]. Note that the many-body contributions
in CCSD(T)/CBS[MP2] are identical to those obtained at RI-MNP2/aug-cc-pVDZ level
(see Section ``System and methods''). The deficiency of correlation-level non-additivity
in MP2 is attributed to the fact that MP2 can take into account only the two-electron
excitation process. In order to describe the correlation-level non-additivity in a system
consisting of four subsystems, MP2 level of theory should be adopted at least,
where four one-electron exciations simultaneously appear in each of the subsystems.~\cite{1994CHA,2000CHA}
Therefore, the CCSD(T) method with the practical approximation considered in the literature~\cite{2006SPO}
can be regarded as being almost same as HF and thus, poor at capturing the
correlation-level non-additivity. Unfortunately, a {\it true} behavior of
the non-additivity at CCSD(T)/CBS level is still unknown.
\par
A comparison between B3LYP and B3LYP-D3 gives the most suggestive insight into
the non-additivity within the framework of DFT.
While the empirical dispersion correction D3 significantly improves the stacking itself
(Fig.~\ref{fig:stacking} (b)), it hardly modifies its non-additivity from B3LYP
(Fig.~\ref{fig:nonadditive}(b)).
This can be attributed to the fact that dispersion corrections based on D3 or
the likes of vdW-XC are additionally made on the original DFT/SCF energies
and thus, never deform their wavefunctions.
From the viewpoint of many-body theory, the deformation of wavefunctions
is the origin of dispersion interactions and hence essential to
the non-additivity.
\par
\subsection{Sign alternation of $\Delta \varepsilon^{(4)}$ in FNDMC}

\vspace{2mm}
Except for B3LYP(-D3) and M06-2X for AA:TT and TA:TA, 
most of the DFT functionals give positive values of
$\Delta \varepsilon^{(4)}$, similar to HF.
We then may conclude that the SCF-level non-additivity is mostly positive definite
(see the previous subsection).
{\it i.e.}, CCSD(T)/CBS[MP2] gives a negligibly small
dispersion-level non-additivity.
In contrast, FNDMC values of $\Delta \varepsilon^{(4)}$ alternate
their signs depending on the corresponding base-pair steps.
This means that the dispersion-level non-additivity
in FNDMC for some base-pair steps
is large enough to change the signs from the SCF-level non-additivity.
According to a simple model analysis based on the London theory (see Introduction),
the dispersion contribution to non-additivity was found to be negative definite.
Thus, we may conclude that the negative non-additivity in FNDMC for some base-pair steps
(AA:TT, TG:CA, and AC:GT) can be attributed to the dispersion contribution.
In summary, the dispersion contributions to non-additivity
have the potential for changing the sign of non-additivity,
depending on base-pair steps.

\vspace{2mm}
If we accept the negative non-additivity in FNDMC in accordance with the London theory,
then another doubt arises about why FNDMC also gives more positive non-additivity,
depending on the base-pair steps.
Comparing FNDMC with HF, FNDMC was found to give
almost the same $\Delta \varepsilon^{(4)}$ as HF
(within errobar) for GG:CC, GC:GC, and CG:CG (Fig.~\ref{fig:nonadditive}).
In contrast, the more positive $\Delta \varepsilon^{(4)}$ deviating from 
the SCF-level non-additivity appears in TA:TA, GA:TC, and AG:CT,
which will be then investigated from the viewpoint of stacking energies.

\begin{figure*}[hbtp]
\begin{center}
  \includegraphics[scale=0.43]{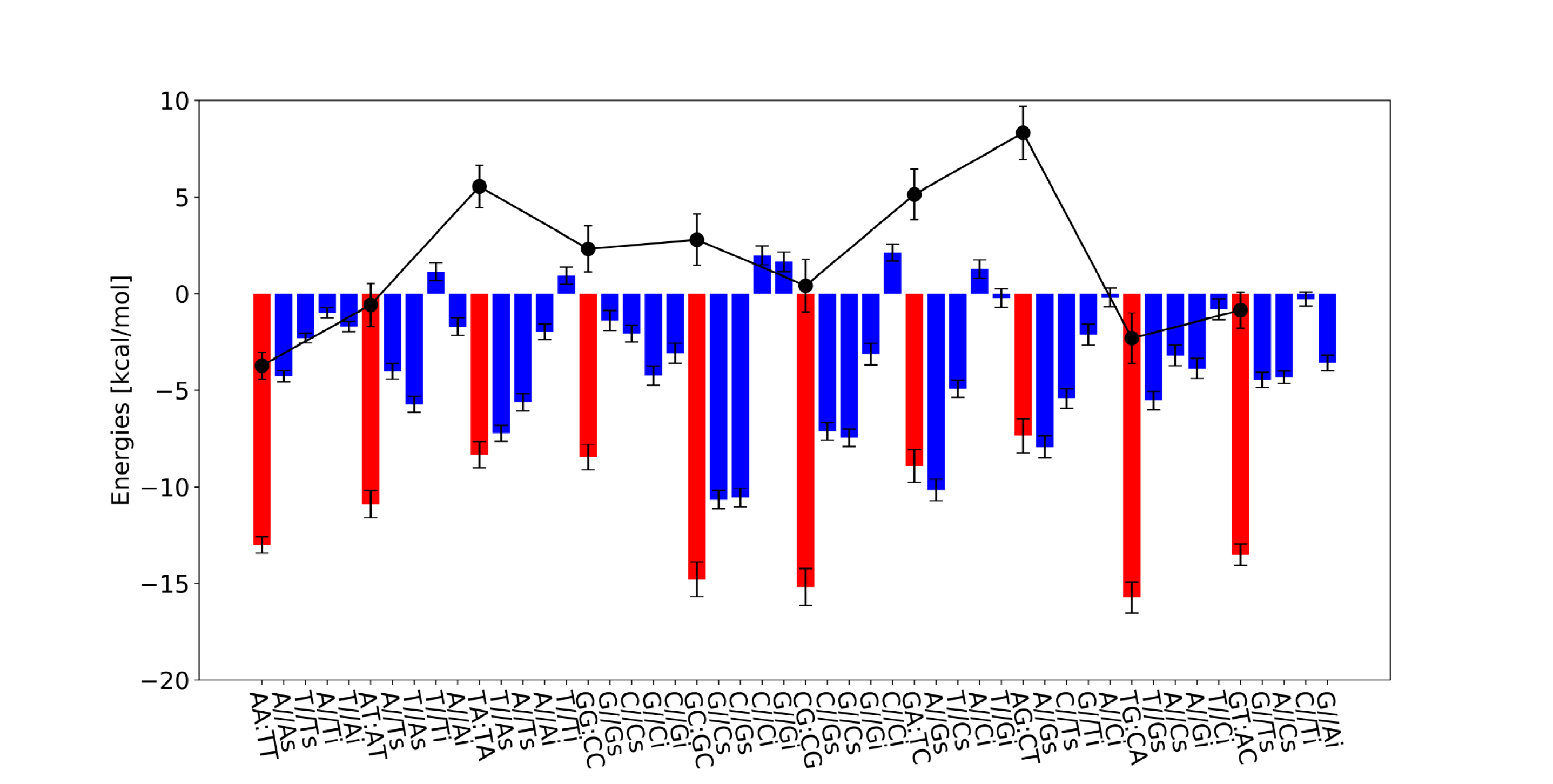}
\end{center}
~\caption{\label{fig:decomposeDMC}
Non-additive contributions (black points) 
decomposed into 4-body (red bars) 
and 2-body (blue bars)stacking energies evaluated 
by DMC [kcal/mol]. 
's' and 'i' ({\it e.g.}, A//Ai) appearing in the labels 
for the horizontal axis indicate intra- and interstrand stacking.
}
\end{figure*}

\vspace{2mm}
Fig.~\ref{fig:decomposeDMC} shows one four-body and 
four two-body stacking energies given in Eq.~(\ref{nonadditive})
for ten unique B-DNA base-pair steps.
It is evident that positive non-additivity
correlates with weaker four-body stacking (red bar),
which is also noted in Fig.~\ref{fig:stacking}
that the base-pair steps with the positive non-additivity
exhibit the weaker stacking described by FNDMC 
than by the other {\it ab initio} methods. 

\begin{figure}[!hbtp]
\begin{center}
  \includegraphics[scale=0.5]{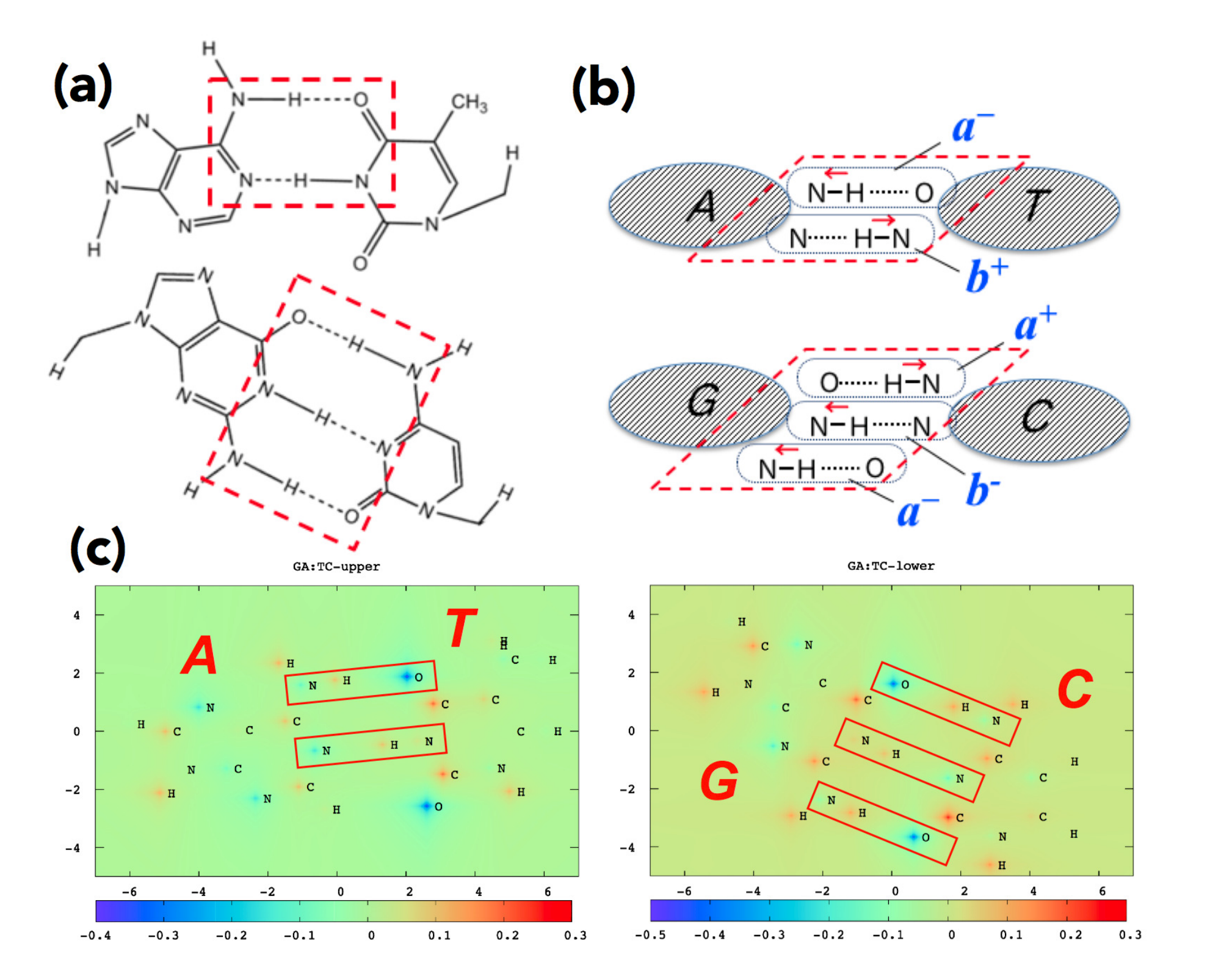}
\end{center}
~\caption{\label{fig:hydLabel}
Hydrogen bondings for GA:TC base pair, 
shown inside the red broken lines [left panel(a)], 
and its schematic picture [panel(b)]. 
Small red arrows put on the N-H bonding 
in the right panel mean the charge transfer 
due to the negativity. 
Bridging bonds can be sorted into 
$a$ (N-H...O) or $b$ (N-H...N), 
and further labelled such as 
$a^+$, $b^-$ {\it etc.}, based on the 
direction of the charge transfer.
Panel(c) shows the Mulliken charge analysis 
for the upper and lower layers. 
Blue and red indicate the negative and positive charge values,
respectively.
}
\end{figure}

\vspace{2 mm}
The weaker four-body staking is identified as
being the origin of the positive non-additivity.
Furthermore, it was found from Fig.~\ref{fig:hydLabel} (a) and (b) 
that the weaker four-body stacking 
can be caused by the bridging bond
between the Watson-Crick base pair:
Although the bridging formed by the hydrogen bonding
partly leads to a stronger stacking in the horizontal direction~\cite{jeffrey2012hydrogen,gilli1989evidence},
it simultaneously weakens the staking 
owing to the vertical repulsion between 
the bridges at different layers (given in SI in detail).
Both the contributions would cancel each other out
and thus, giving rise to the overall 'weaker stacking'.
This cancellation causes the positive non-additivity 
depending on the base-pair steps.
This factor was not taken into
account in our simple London model analysis.

\vspace{2 mm}
To estimate the contributions quantitatively, 
we first evaluated the Mulliken charge 
that appeared in the bridging location, 
as shown in Fig.~\ref{fig:hydLabel} (c). 
Based on the charge, we then evaluated 
the Madelung repulsion interaction, 
getting +5$\sim$10 kcal/mol per bond. 
Since the typical range of the 
energy gain due to the hydrogen bonding 
is known to be less than 5$\sim$6kcal/mol~\cite{mcnaught1997compendium},
it is likely to result in a 
positive contribution to non-additivity,
thus making the stacking weaker.

\section{Discussion}
The present results are unexpected against
a common consensus among the methodologies, 
and hence likely to mislead readers.
Here we shall clarify what points are to be noted
in order to avoid misleading.
The first point is that the present results 
never immediately lead to such a simple 
controversy as 'DMC vs CCSD(T)'.
The second point is that the 
unexpected results are obtained 
only at the non-additivity level (Fig.~\ref{fig:nonadditive}), 
but not at the stacking energy level (Fig.~\ref{fig:stacking}). 

\vspace{2mm}
As for the first point, our results in Fig.~\ref{fig:nonadditive} 
are {\it apparently} regarded as
'DMC vs [CCSD(T) together with all the other DFT/SCF methods]'. 
This might be followed by such a misleading discussion as 
'which method is reliable?', but it is actually premature 
to draw a conclusion about this question. 
We should clarify limitations on approximations 
adopted in the present FNDMC and CCSD(T) simulations.

\vspace{2mm}
We start with CCSD(T) together with HF/DFT.
It was seen that HF and B3LYP failed to
describe the B-DNA stacking interactions
(Fig.~\ref{fig:stacking} (a) and (b)),
but their non-additive contributions were almost same
as ``CCSD(T)'' (Fig.~\ref{fig:nonadditive} (a) and (b)).
This must be unacceptable because HF and B3LYP
do not include dispersion interactions
at all.~\cite{1994CHA,2000CHA,2010GRI}
Furthermore, all the other DFT functionals give
almost the same trend as CCSD(T)/CBS[MP2],
though each of them slightly deviates from CCSD(T)/CBS[MP2].
In spite of their failure in reproducing the staking,
the HF and B3LYP methods demonstrate almost the same 
$\Delta \varepsilon^{(4)}$ as the other DFT methods.
In particular, the dispersion correction based on B3LYP-D3
significantly improves the staking itself, but 
seldom changes the non-additive contributions from B3LYP.
This means that reproducibility of $\varepsilon^{(4)}$ hardly
affect that of $\Delta \varepsilon^{(4)}$ at SCF level.
MP2 can reproduce a main part of the dispersion interaction
(or electron correlation) by the second-order perturbation
energy correction,
but its wavefunction still remains the Hartree-Fock one.
This is analogous with the relation B3LYP and B3LYP-D3.
Thus, we may conclude that the dispersion correction 
accompanied with wavefunction deformation is essential
for reproducing the dispersion-level non-additivity. 

\vspace{2mm}
The overall coincidence between ``CCSD(T)''
and SCF (HF/DFT) implies that 
(1) the present ``CCSD(T)'' method never describe
the dispersion-level non-additivity and thus,
losing a large part of {\it true} non-additivity
at CCSD(T) level of theory, or (2) 
a {\it true} dispersion-level non-additivity
is essentially tiny and thus, well described
by the present ``CCSD(T)'' method.
As has been explained in ``System and methods'',
the present ``CCSD(T)'' method relies on
the CBS[MP2] approximation, {\it i.e.},
it is not a {\it true} ``CCSD(T)/CBS''.
Since the true CCSD(T)/CBS is believed
to well reproduce the non-additivity,
it is unlikely for CCSD(T)/CBS to be
insufficient to describe the non-additivity.
So the second statement is implausible,
while the first one is to be studied in more detail.
Hereafter we investigate whether or not 
the CBS[MP2] approximation can be
a possible source of damaging
a capability inherent in ``CCSD(T)/CBS''
of capturing the non-additivity.

\begin{figure}[!hbtp]
\begin{center}
  \includegraphics[scale=0.75]{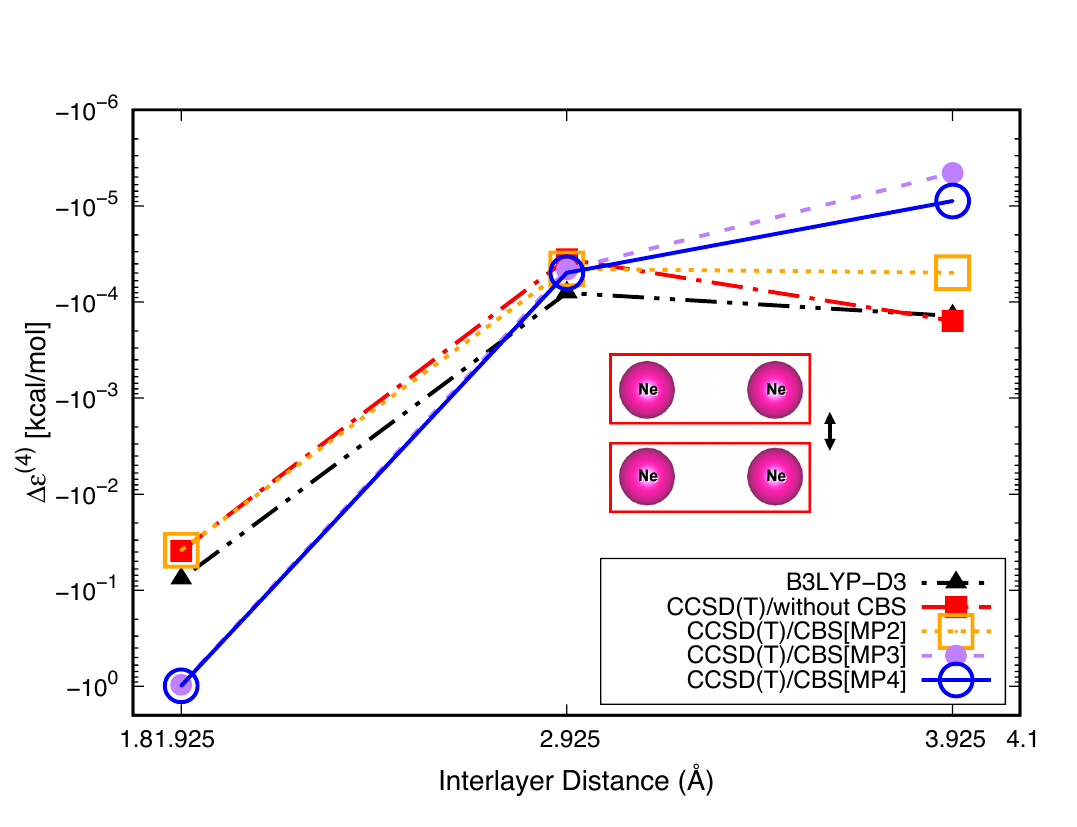}  
\end{center}
~\caption{\label{fig:NeQuadramer}
  The non-additivity $\Delta \varepsilon^{(4)}$ of
  neon tetramer at several distances between the constituent dimers
  (described as ``Interlayer distance'' in the horizontal axis).
  With a fixed ``Interlayer distance'',
  all the Ne atoms located on a plane form a rectangle,
  where in each dimer its interatomic length is
  fixed to be 2.925 \AA.
  All the CCSD(T) and DFT calculations were performed
  using Gaussian09~\cite{2009FRI}.
}
\end{figure}

\vspace{2mm}
To address the above issue, we attempted to apply
CBS[MP3] and CBS[MP4] (as well as the true CBS)
to the B-DNA base-pair steps, 
but they were too large to compute.
Instead, we dealt with
a neon tetramer as a simple/model system
in which each DNA base is replaced by a Ne atom.
We evaluated $\Delta \varepsilon^{(4)}$ values of the Ne tetramer
at several distances between the two dimers
using CCSD(T)/VTZ (w.o. CBS), CCSD(T)/CBS[MP$n$] ($n = 2, 3, 4$),
and B3LYP-D3.
Fig.~\ref{fig:NeQuadramer} shows how those values
differ from each other:
CCSD(T)/CBS[MP2] is significantly different
from CCSD(T)/CBS[MP3] and CCSD(T)/CBS[MP4],
but it is almost same as B3LYP-D3.
This indicates that the CBS[MP2] level
can be regarded as the SCF-level non-additivity,
which is consistent with our finding in the B-DNA case
that CCSD(T)/CBS[MP2] is incapable of
reproducing the dispersion-level non-additivity properly.
In contrast, CCSD(T)/CBS[MP3] almost converges to CCSD(T)/CBS[MP4],
while it significantly deviates 
from both CCSD(T)/CBS[MP2] and B3LYP-D3 -- SCF-level non-additivity.
From the viewpoint of the perturbation theory,
this convergence means that a main contribution
to the dispersion-level non-additivity
can be described by the CBS[MP3] level.
That is, at least CBS[MP3] is essential
to describe the dispersion-level non-additivity,
while CBS[MP2] is insufficient.
Although we could not actually confirm how
CBS[MP3/4] differ from SCF/CBS[MP2] for the B-DNA case,
we may infer that even in the B-DNA case
CBS[MP3/4] dominantly contribute to
a description of the dispersion-level non-additivity.

\vspace{2mm}
Next we move on to FNDMC.
Its wiggling dependence of $\Delta \varepsilon^{(4)}$
on the base-pair steps appearing in Fig.~\ref{fig:nonadditive}
arouses suspicion whether FNDMC really reproduces
the dispersion-level non-additivity appropriately.
Here we shall deliberate on two possibilities of
causing faults in FNDMC: quality of
trial wavefunctions obtained and reliability of the fixed-node
approximation adopted in the present study.

\vspace{2mm}
We first note that such a wiggling dependence
appears only in Fig.~\ref{fig:nonadditive}, 
but not in Fig.~\ref{fig:stacking};
both the results were obtained by the wavefunctions
that were optimized at the same level of theory.
In the present study we did not optimize
the Slater part the trial wavefunctions,
but the Jastrow part only.
In FNDMC, the latter changes the statistical error bar only,
while the former -- related to the fixed-node approximation --
changes the final total energy value,
unlike VMC (variational Monte Carlo).~\cite{1998KWO}
It is well known that the same performance on
the Jastrow optimization leads to
the same magnitude of error bars for similar system sizes
if all the other computational details are assumed to be common
to the systems.
It is obvious from
Figs.~\ref{fig:stacking} and~\ref{fig:nonadditive}
that this is actually valid for the present B-DNA systems.
We also insist that our choice of computational details on FNDMC
-- basis sets, time step, t-move scheme,
as well as Jastrow function -- 
is equivalent to a protocol established in previous studies
on non-covalent systems 
due to Dubeck\'{y} {\it et al.}~\cite{2014DUB}.
The reasonable behavior of FNDMC stacking energies
in Fig.~\ref{fig:stacking} asserts that our choice is
valid for the present B-DNA base-pair steps.

\vspace{2mm}
The fixed-node approximation is the most notorious as
the cause of errors in FNDMC.~\cite{1982REY}
Previous studies on non-covalent systems including B-DNA,
however, demonstrated that FNDMC works well for 
evaluating their complexation energies
in general.~\cite{2014DUB}
It is to be noted in the B-DNA stacking
that this is valid for the stacking energies,
but unknown for the non-additivity.
The success in the FNDMC stacking energies
relies on the error {\it cancellation} of
the fixed-node approximations between
the whole non-covalent system and its constituent sub-systems.
This implies that the formation of non-covalent/vdW bonding
does not give rise to a significant difference
in nodal surface structures
between the whole and the sub-systems
(tetramer-dimer/dimer-monomer),
leading to an accurate complexation energy.
The success in the non-additive contribution
requires that two error cancellations
of the fixed-node approximations simultaneously
occur for $\varepsilon^{(4)}$ and $\varepsilon^{(2)}$.
In the case of non-additivity, however, 
it is possible that 
the cancellation in $\varepsilon^{(4)}$
would not occur properly.
Its possible factor can be attributed
to a horizontal bridging between Watson-Crick bases
due to hydrogen bonding.
The formation of hydrogen bonding accompanied by
the charge transfer could deform the fixed-node surface
structure of the tetramer more significantly than
that of vdW bonding.
If this were true, the fixed-node error 
could not be canceled out more remarkably
in $\varepsilon^{(4)}$ than in $\varepsilon^{(2)}$.
The less cancellation could arouse the suspicion that
the wiggling dependence of FNDMC in Fig.~\ref{fig:nonadditive}
is incorrect due to the fixed-node error
related to the hydrogen bonding.

\vspace{2mm}
In order to prove the conjecture (or anti-conjecture) about 
the fixed-node errors caused by the hydrogen bonding, 
the most straightforward way would be 
to evaluate the nodal surface 
dependence of the non-additivity. 
While the stacking energy has been
demonstrated to be insensitive to
the dependence~\cite{2013HON,2016HON},
one would suspect that it is 
not the case for the non-additivity. 
Suppose it were true, 
the non-additivity evaluated with a different trial node 
would be different from the present FNDMC one 
in Fig.~\ref{fig:nonadditive}. 
Although we plan to address this issue 
in our future work, we should mention that
such a calculation involves a heavy computational resource,
which costs $1.2 \times 10^6$ core-hour 
[$1.2 \times 10^5$ (core-hour) $\times$ 10 pairs]. 
In addition to a single reference trial node,
we could employ recently developed trial nodes such 
selected configuration interaction~\cite{2018GAR}
and multipfaffian~\cite{2008BAJ} wavefunctions,
as well as a simple multi-reference
one~\cite{2012HON,2017ICH}.
Beside their feasibility in terms of computation costs,
the more sophisticated trial nodes would shed light on
the nodal surface dependence of non-additivity in FNDMC,
{\it i.e.}, we could verify whether or not the non-additivity
is sensitive to the nodal surface 
unlike the stacking energies~\cite{2013HON,2014DUB}.

\begin{figure}[!hbtp]
\begin{center}
  \includegraphics[scale=0.8]{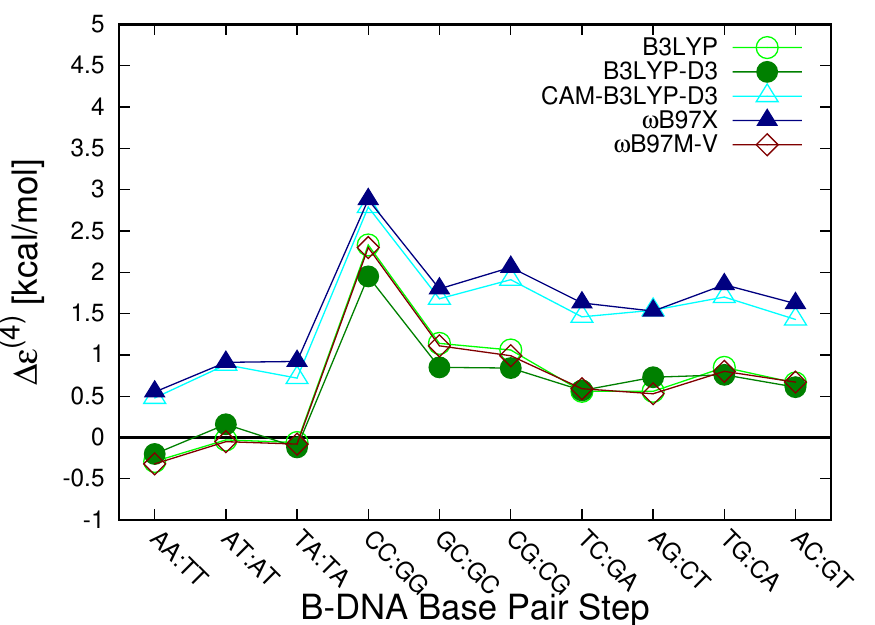}  
\end{center}
~\caption{\label{fig:DFTchargeTransfer}
    Non-additive contributions, $\Delta E^{(4)}$ [kcal/mol], 
    predicted by different XC functionals with/without the
    long-range exchange corrections.
    The charge transfer mainly occurs at horizontal
    hydrogen bonding when forming Watson-Crick bases.
}
\end{figure}

\vspace{2mm}
Although we do not investigate
the nodal surface dependence of $\Delta \varepsilon^{(4)}$ further,
we alternatively examine if the charge transfer
-- the key to verifying the issue -- could 
really matter even for the SCF-level non-additivity.
Suppose it really matters, one would expect that
the charge transfer could somewhat affect 
the dispersion-level non-additivity.
We consider two types of XC functionals in terms of
the long-range (LC) exchange corrections: 
one well describes the charge transfer with the correction
and the other dose not.
Their difference in the charge density distribution tells us 
how significantly the distribution in a B-DNA base-pair
change before and after forming the base-pair,
thus clarifying an effect of changing 
their non-additive contributions.
For example, it is well known that B3LYP-D3
is not good at capturing the charge transfer
because B3LYP also fails for a number of cases
relevant to the charge transfer~\cite{2004YAN}
and the D3 correction never improve the
B3LYP description of charge density~\cite{2010GRI}.
On the other hand, CAM-B3LYP-D3~\cite{2004YAN}
remarkably improve the charge transfer,
because it enhances the exact exchange for the long-range exchange
based on Coulomb-Attenuating Method (CAM).
As another choice, $\omega$B97X has been reported 
to give a better descriptions of properties
including the charge transfer than $\omega$B97M-V
in some cases.~\cite{2011SIN}
We note that there are further choices of XC
for reproducing the charge transfer well,
such as 'self-consistent vdW'
implemented in a series of 'vdW-DF'~\cite{2015BER},
but their implementations are unavailable for
Gaussian basis set calculations.
Fig.~\ref{fig:DFTchargeTransfer} focuses on
a comparison among XC functionals with/without
long-range corrections
(originally taken from Fig.~\ref{nonadditive})  
We found that the long-range corrections by
CAM-B3LYP(-D3) and $\omega$B97X (positively) enhance
the non-additive contributions compared to
the counterparts, B3LYP(-D) and $\omega$B97M-V.
This implies the importance of the charge transfer
caused by the hydrogen bonding when forming
a Watson-Crick base pair.
Although the above analysis deals with
only the SCF-level non-additivity,
it would be expected that
the charge transfer caused by the hydrogen bonding
could significantly deform the nodal surface structures
when forming the Watson-Crick base 
giving rise to a large fixed-node error,
and hence the wiggling dependence of $\Delta \varepsilon^{(4)}$
in FNDMC, shown in Fig.~\ref{fig:nonadditive},
might be false due to the less error cancellation.
in the $\Delta \varepsilon^{(4)}$ evaluation.

\vspace{2mm}
Lastly, we mention another drawback 
for FNDMC simulations in a practical sense.
According to the previous analysis on the Ne tetramer,
it might be expected that
CCSD(T)/CBS[MP$n$] ($n \ge 3$) would 
get closer to FNDMC predictions 
as increasing the CBS[MP$n$] level.
For comparison, we attempted to
apply FNDMC to evaluate $\Delta \varepsilon^{(4)}$ of
the Ne tetramer. Unfortunately, however, we could not
obtain numerically/statistically reliable FNDMC results
because the magnitude of $\Delta \varepsilon^{(4)}$ itself is
an order of/less than the sub-chemical accuracy (0.1 kcal/mol)
and hence the corresponding error bar is required to be
an order of/less than 0.01 kcal/mol.
To attain such an error bar, 
a vast number of statistical samplings 
must be accumulated even for the smaller system considered here.
This is another serious drawback to be noted for FNDMC.

\section{Conclusion}
We applied a diffusion Monte Carlo method
with the fixed-node approximation (FNDMC) 
to evaluate the non-additive contribution to 
B-DNA stacking.
We found that the FNDMC values of non-additivity
alter their sign ({\it i.e.} they increase or decrease
their stacking interactions)
depending on the base-pair steps,
which is contrary to all the other {\it ab initio} methods.
On the other hand, no significant difference between
the methodologies was observed for 
four-/two-body stacking energies, each of which
are used to evaluate the non-additivity.
To elucidate this contrast
between the stacking and non-additivity, 
we made two plausible discussions about limitation on
practical approximations involved in CCSD(T) and FNDMC:
(1) The reason why the unexpected coincidence 
between CCSD(T)/CBS[MP2] and HF/B3LYP
occurs only at the non-additivity level
can be attributed to 
the imperfect capability for MP2 
to reproduce the electron correlation 
specific to the four-body system.
The lack of the correlation
never describes the correlation/dispersion-level
non-additivity properly.
In other words, CCSD(T)/CBS[MP2] mostly describes the
SCF-level non-additivity only.
(2) FNDMC demonstrates
a wiggling dependence of the non-additivity.
While the SCF-level non-additivity is mostly positive,
the non-additive contributions described by FNDMC
are both positive and negative signs.
The negative sign is found to be reasonable,
which might be supported by a simple model analysis
based on the London theory.
It would, however, be premature 
to draw a conclusion that the FNDMC non-additivity
reveals the truth.
This is because the Watson-Crick base-pair
involves the charge transfer caused by the hydrogen bonding,
but we could not verify if the error cancellations of
the fixed-node errors were successful for the hydrogen bonding,
as in the case of complexation energies.
In the near future, we will investigate
the fixed-node dependence of the non-additivity
by considering more reliable trial nodal surfaces,
albeit with huge computational costs.

\section*{Acknowledgments}
The computation in this work has been performed 
using the facilities of the Research Center for Advanced
Computing Infrastructure (RCACI) at JAIST.
T.I. is grateful for financial suport from Grant-in-Aid
for JSPS Research Fellow (18J12653).
K.H. is grateful for financial support from
a KAKENHI grant (JP17K17762), Grant-in-Aid for
Scientific Research on Innovative Areas
``Mixed Anion'' project (JP16H06439) from MEXT, 
PRESTO (JPMJPR16NA) and the Materials research by
Information Integration Initiative (MI$^2$I) project 
of the Support Program for Starting Up Innovation Hub
from Japan Science and Technology Agency (JST). 
R.M. is grateful for financial supports from MEXT-KAKENHI (17H05478 and 16KK0097), 
from Toyota Motor Corporation, from I-O DATA Foundation, 
and from the Air Force Office of Scientific Research (AFOSR-AOARD/FA2386-17-1-4049).
R.M. and K.H. are also grateful to financial supports
from MEXT-FLAGSHIP2020 (hp170269, hp170220).

\bibliography{references}


\end{document}